\documentclass[12pt]{iopart}
\usepackage{graphicx}

\begin{document}

\title[Energy-momentum tensor form factors of the nucleon]{
Energy-momentum tensor form factors of the nucleon within 
a $\pi$-$\rho$-$\omega$ soliton model}

\author{Ju-Hyun Jung$^1$, Ulugbek Yakhshiev$^2$ and
Hyun-Chul Kim$^3$}

\address{Department of Physics, Inha University, 402-751 Incheon,
  Republic of Korea}
\ead{$^1$juhyun@inha.edu, $^2$yakhshiev@inha.ac.kr,
  $^3$hchkim@inha.ac.kr} 
\begin{abstract}
We investigate the energy-momentum tensor form factors of the nucleon
within the framework of a chiral soliton model, including the $\rho$
and $\omega$ vector mesons. We examine the role of each meson degrees
of freedom in these form factors. It is explicitly shown that the pion
provides strong attraction whereas the $\rho$ and $\omega$ yield
repulsion in such a way that the soliton becomes stabilized.  
The results are discussed in comparison with those of other models. 
\end{abstract}

\pacs{14.20.Dh, 21.65.Jk}

\section{Introduction}
Nucleon form factors are essential quantities in understanding the
internal structure of the nucleon. For example, the electromagnetic
form factors reveal how the charge and the magnetization of quarks are
distributed inside a nucleon. The scalar and axial-vector form factors
also provide information on certain aspects of the nucleon structure
such as chiral and flavor symmetries and breakdown of them. Because of
these reasons, a great deal of investigations has been performed
extensively over decades. On the other hand, the energy-momentum
tensor form factors (EMTFFs) of the nucleon, even though they were
proposed in 1966~\cite{Pagels:1966zza}, has attracted attention only
very recently, since there is no probe to measure them
directly. However, the EMTFFs are as equally important as the
electromagnetic FFs, since they also provide crucial information on
how the internal structure of the nucleon. The generalized parton
distributions (GPDs) enable one to extract them from hard exclusive 
reactions~\cite{Muller:1998fv,Ji:1996nm,Collins:1996fb,
  Radyushkin:1997ki}.     
The Melin transforms of certain GPDs can be identified as the EMTFFs
that expose how the mass and the spin are distributed inside a 
nucleon. Moreover, the EMTFFs can be regarded as the touchstone of  
checking the validity of any model for the nucleon: they provide 
strong constraints on the model in such a way that the pressure should
be zero. Moreover,the $D$-term, one of the EMTFFs, was deeply related
to the spontaneous breakdown of chiral symmetry~\cite{Polyakov:1999gs,
  Kivel:2000fg, Goeke:2001tz}. Thus, the EMTFFs give us a whole new
perspective on the structure of the nucleon. 

The EMTFFs are defined as a nucleon matrix element of the totally
symmetric Energy-momentum tensor (EMT) operator as 
follows~\cite{Ji:1996ek,Polyakov:2002yz}:
\begin{eqnarray}
  \langle p^\prime| \hat T_{\mu\nu} (0) |p\rangle
 & =& \bar u(p^\prime,\,s') \left[M_2(t)\,\frac{P_\mu P_\nu}{M_N}+
  J(t)\ \frac{i(P_{\mu}\sigma_{\nu\rho}+P_{\nu}\sigma_{\mu\rho})
    \Delta^\rho}{2M_N}\right.\cr
&&\left.  + d_1(t)\,
  \frac{\Delta_\mu\Delta_\nu-g_{\mu\nu}\Delta^2}{5M_N}\right] u(p,\,s)\, ,
  \label{Eq:EMTff}
\end{eqnarray}
where $P=(p+p')/2$, $\Delta=(p'-p)$ and $t=\Delta^2$.
The $M_N$ and $u(p,\,s)$ denote the nucleon mass and spinor,
respectively. The form factor $M_2(t)$ gives information about the
ratio of the momenta carried by constituents of a nucleon. In
particular, $M_2(t)$ at the zero-momentum transfer shows that 
about 1/2 of the momentum of a fast moving nucleon is carried by
quarks, and the other half by gluons.  The other form factor $J(t)$
reveals information on the total angular momentum of the quark and
gluons, though it is not much known experimentally. It is less trivial
to understand the physical meaning of the last form factor $d_1(t)$ in
Eq.(\ref{Eq:EMTff}) but is equally important, since it explains how
the strong forces are distributed and stabilized in the
nucleon~\cite{Polyakov:2002yz,Polyakov:2002wz}. It can be extracted
from the beam charge asymmetry in deeply virtual Compton
scattering~\cite{Kivel:2000fg}.    

The EMT form factors of the nucleon have been investigated
in various approaches, for example,  
in lattice
QCD~\cite{Mathur:1999uf,Hagler:2003jd,Gockeler:2003jf, Negele:2004iu,
Brommel:2007sb, Bratt:2010jn, Liu:2012nz,Hagler:2007xi},
in chiral perturbation theory
\cite{Chen:2001pv, Belitsky:2002jp,Ando:2006sk,Diehl:2006ya,Dorati:2007bk},
in the chiral quark-soliton model ($\chi$QSM)
\cite{Petrov:1998kf, Schweitzer:2002nm, Ossmann:2004bp,
  Wakamatsu:2005vk, Wakamatsu:2006dy, Wakamatsu:2007uc, Goeke:2007fp,
  Goeke:2007fq} as well as in the Skyrme model~\cite{Cebulla:2007ei}. 
Those of nuclei have also been studied
\cite{Polyakov:2002yz, Guzey:2005ba, Liuti:2005gi, Scopetta:2009sn}.
The study of the nucleon EMTFFs was also extended to nuclear
matter~\cite{Kim:2012ts}. In the present work, we want to examine the 
EMTFFs of the nucleon, based on a chiral soliton model with vector
mesons in the minimal form~\cite{Meissner:1986ka}.\footnote{In the
  present work, we consider one of the simplest ones among various
  soliton models~\cite{Meissner:1986ka,
    Igarashi:1985et,Schwesinger:1986xv,Jain:1987sz} with vector
  mesons. For the detailed development of chiral solitons with vector
  mesons, we refer to a recent review~\cite{Weigel}.}  The 
model is based on the fact that the nonlinear sigma model has a hidden
local gauge symmetry $\mathrm{SU}(2)_V$~\cite{Bando:1984ej}, in which
the $\rho$ meson is identified as a gauge boson. Extending this
symmetry to $\mathrm{SU}(2)_V\otimes \mathrm{U}(1)$, the $\omega$
meson can be also regarded as a gauge boson~\cite{Meissner:1986ka}. In
this way, parameters of the model are completely determined in the
mesonic sector, so that we can investigate properties of the nucleon
in the solitonic sector unambiguously. This model has a certain virtue
in studying the EMTFFs, since one can study the role of the $\rho$ and 
$\omega$ mesons in describing the nucleon. It is known that the vector
mesons provide short-range repulsion in the one-boson exchange model
for the nucleon-nucleon interaction, while the pion dominates the
long-range interaction~\cite{Machleidt:1987hj}. We will soon show
explicitly that the pion gives attraction to form a soliton whereas
the vector mesons become repulsive to stabilize it, which is analogous
to the nucleon-nucleon interaction. This is in line with what was
found in the Skyrme model~\cite{Cebulla:2007ei}, where it was shown 
for the Skyrme term to stabilize obviously the soliton. 

The present work is organized as follows: In Section II, we briefly
introduce the chiral soliton model with the $\rho$ and $\omega$
mesons. In Section III, we explain how to compute the EMTFFs of the
nucleon within this model. In Section IV, we discuss the results of
the EMTFFs. The final Section is devoted to summary and conclusion.

\section{Chiral soliton model with vector mesons}

We start with the effective Lagrangian with the $\pi$, $\rho$, 
and $\omega$ meson degrees of freedom, from which the nucleon arises
as a topological 
solution~\cite{Meissner:1986ka,Meissner:1986js,Meissner:1987ge}. In 
fact, almost all formulae presented in this section can be found in  
Ref.~\cite{Meissner:1987ge}. We briefly recapitulate here the
pertinent ones for the EMTffs. 
The Lagrangian has the following form
\begin{eqnarray}
\mathcal{L} & = & 
\mathcal{L}_\pi
+\mathcal{L}_{\rm kin}+\mathcal{L}_V+
\mathcal{L}_{\rm WZ}\,,
\label{lag}\\
\mathcal{L}_\pi&=&\frac{f_{\pi}^{2}}{4}{\rm Tr}\left(\partial_{\mu}
  U\partial^{\mu}U^{\dagger}\right) + 
 \frac{f_{\pi}^{2}m_{\pi}^{2}}{2}{\rm Tr}\left(U-1\right)\,,\\
\mathcal{L}_{\rm kin}&=&
-\frac{1}{2g^{2}}{\rm Tr}\left(\partial_\mu V_\nu-\partial_\nu
V_\mu-i[V_\mu,V_\nu]\right)^2\,,\\
\mathcal{L}_V&=&
\frac{a}{4}f_{\pi}^{2}{\rm Tr}\left[D_{\mu}\xi\cdot\xi^{\dagger}
  +D_{\mu}\xi^{\dagger}\cdot\xi\right]^{2} \,,\\ 
\mathcal{L}_{\rm WZ}&=&
 \left(\frac{N_{c}}{2}g\right)\omega_{\mu}
 \frac{\epsilon^{\mu\nu\alpha\beta}}{24\pi^{2}}{\rm Tr}\left\{
   \left(U^{\dagger}\partial_{\nu}U\right)
   \left(U^{\dagger}\partial_{\alpha}U\right)
   \left(U^{\dagger}\partial_{\beta}U\right)\right\},   
\end{eqnarray}
where $U = \xi_{L}^{\dagger}\,\xi_{R}$ in unitary gauge, and the
covariant derivative is defined as
\begin{eqnarray}
D_{\mu}\,\xi_{L(R)}& = & \partial_{\mu}\,\xi_{L(R)}-i\,
V_{\mu}\,\xi_{L(R)}\,. 
\end{eqnarray}
The field $V_\mu$ consists of the $\rho$ and $\omega$
fields, i.e. $\vec{\rho}_\mu$ and $\omega_\mu$, respectively, being 
expressed as  
\begin{equation}
  \label{eq:vectorf}
 V_\mu=\frac{g}2\,(\vec\tau\cdot\vec\rho_\mu
+\omega_\mu).  
\end{equation}
The pion decay constant $f_\pi$ and the pion mass $m_\pi$, are
fixed by experimental data, i.e. $f_\pi=93$ MeV and $m_\pi=135$ MeV
(the neutral pion mass). The number of colors is taken to be $N_c=3$
and the coupling constant $g$ is related to the
Kawarabayashi-Suzuki-Riazuddin-Fayyazuddin (KSRF)
relation~\cite{Kawarabayashi:1966kd, Riazuddin:1966sw} 
$m_{\rho}^{2} = m_{\omega}^{2} =  a\, g^{2}\, f_{\pi}^{2}$ with
$a=2$ such that we have the $\rho\pi\pi$ coupling
$g_{\rho\pi\pi}={ag}/{2}$ and $g=5.85$. Note that the $g_{\rho\pi\pi}$ 
is taken to be close to its empirical value $g_{\rho\pi\pi}=6.11$.  

Assuming the following Ans\"atze for the pseudoscalar and vector mesons  
\begin{equation}
  \label{eq:ansatz}
U \;=\; \exp\left\{\frac{i\vec\tau\cdot\vec r}{r} F(r)\right\}\,,\quad 
\rho_\mu^a=\frac{\varepsilon_{ika}r_k}{gr^2}\,G(r)\delta_{\mu i}\,, 
\quad \omega_\mu=\omega(r)\delta_{\mu0}\,,  
\end{equation}
one can derive the static energy functional from the Lagrangian. It is
identified as the classical soliton mass 
\begin{eqnarray}
M_{\mathrm{sol}}&=& 4\pi\int_0^\infty
\mathrm{d}r\left\{\frac{f_{\pi}^{2}}{2}\left(r^2F^{\prime\,2}  
 +{2\sin^{2}F}\right)
 +r^2f_{\pi}^{2}m_{\pi}^{2}\left(1-\cos F\right)\right.\cr
 && +\,{2f_{\pi}^{2}}\left(G+1-\cos F\right) ^{2}
 +\,\frac{1}{g^{2}}\left[G^{\prime\,2} +
     \frac{G^2(G+2)^2}{2r^2} \right]\cr
&&\left.-\,r^2\left(f_\pi^2g^{2}\omega^2+\frac12\,\omega^{\prime\,2}\right) 
  + \frac{3g}{4\pi^2}\,\omega F'\sin^2F \right\}, 
\end{eqnarray}
where $f'=\partial f/\partial r$, generically. 
Minimizing the  classical soliton mass can be achieved by solving the
equations of motion, which are given as the coupled nonlinear
differential equations 
\begin{eqnarray}
F^{\prime\prime}&=&-\frac{2}{r}\,
F'+\frac{1}{r^2}\left[4(G+1)\sin F-\sin
  2F\right]  +  m_\pi^2 \sin F-\frac{3g\omega'}{4\pi^2f_\pi^2}\,
\frac{\sin^2F}{r^2}\,,\cr  
G^{\prime\prime}&=&2g^2 f_\pi^2\left[G+2\sin^2\frac{F}2\right] +
\frac{G(G+1)(G+2)}{r^2} 
\nonumber\,,
\qquad\\
\omega^{\prime\prime}&=&-\frac{2}{r}\,\omega' + 2f_\pi^2
g^{2}\omega-\frac{3g}{4\pi^2r^2}F'\sin^2F\, 
\label{statEQ}
\end{eqnarray}
with the boundary conditions
\begin{equation}
  \label{eq:15}
F(0)=\pi,\;\;\; G(0)=-2,\;\;\;
F(\infty)=G(\infty)=\omega(\infty)=\omega'(0)=0\,.
\end{equation}

The collective quantization brings out the following relations
\begin{eqnarray}
U(\vec{r},\,t) &=& A(t)U(\vec{r})A^+(t)\,,\cr 
\omega_i(\vec r,t) &=& \frac{\phi(r)}{r}\, \left(\vec{K}\times
  \frac{\vec{r}}r\right)_i, \cr 
\vec{\tau}\cdot\vec{\rho_0} (\vec{r},\,t) &=& \frac{2}{g} A(t)
\vec{\tau}\cdot\left[\vec{K} \xi_1(r) + \frac{\vec{r}}r
  \left(\vec{K} \cdot 
  \frac{\vec{r}}r\right) \xi_2(r)\right] A^+(t),\cr
\vec{\tau}\cdot\vec{\rho}_i(\vec{r},\,t) &=& A(t) \vec{\tau} \cdot
\vec{\rho}_i(\vec{r}) A^+(t),
\end{eqnarray}
where $2\vec{K}$ denotes the angular velocity of the soliton with   
the relation $i\vec{\tau}\cdot \vec{K} = A^+\dot{A}$. This leads to the
time-dependent collective Hamiltonian
\begin{equation}
H(t) = M_{\mathrm{sol}} +\Lambda{\rm Tr}(\dot A\dot A^+),   
\end{equation}
where $\Lambda$ denotes the moment of inertia of
the rotating soliton 
\begin{eqnarray}
\Lambda & = & 4\pi\int_0^\infty\!\!
dr\left\{\frac{2}{3}f_{\pi}^{2}r^{2}\left(\sin^{2}F+ 
8\sin^{4}\frac{F}{2}-8\xi_{1}\sin^{2}\frac{F}{2}+3\xi_{1}^{2}
+2\xi_{1}\xi_{2}+\xi_{2}^{2}\right)\right.\nonumber\\  
 &  & \qquad\qquad\left.+\frac{1}{3g^{2}}\left[
4G^{2}\left(\xi_{1}^{2}+\xi_{1}\xi_{2}-2\xi_{1}-\xi_{2}
  +1\right)\right.\right.\nonumber\\ 
 &  & \qquad\qquad\left.+2\left(G^{2}+2G+2\right)\xi_{2}^{2} 
+r^{2}\left(3\xi_{1}^{\prime2}+\xi_{2}^{\prime2}+2\xi_{1}'\xi_{2}'
\right)\right]\nonumber \\
 &  &  \qquad\qquad\left.-\frac{1}{6}
   \left[\Phi'^{2}+\frac{2\Phi^{2}}{r^{2}}+  
2\left(gf_{\pi}\right)^{2}\Phi^{2}\right] +g\frac{\Phi
F'}{2\pi^{2}}\sin^{2}F\right\}\,. 
\end{eqnarray}

In the large $N_c$ expansion, one extremizes the moment of inertia
and gets the coupled nonlinear differential equations for the next-order 
profile functions $\xi_1$, $\xi_2$, $\phi$
in the presence of the leading-order profile functions $F$, $G$ and
$\omega$ 
\begin{eqnarray}
\xi_{1}'' & = & 2f_{\pi}^{2}g^{2} \left(\xi_{1}-1+\cos
  F\right)-\frac{2\xi_{1}'}{r} +
\frac{G^{2}\left(\xi_{1}-1\right)+2(G+1)\xi_{2}}{r^{2}}\,,\cr
\xi_{2}'' & = & 2f_{\pi}^{2}g^{2}\left(\xi_{2}+1-\cos F\right)
-\frac{2\xi_{2}'}{r} + \frac{G^{2}\left(\xi_{1}-1\right) + 
2\left(G^2+3G+3\right)\xi_{2}}{r^{2}}\,,\cr
\phi'' & = & 2f_{\pi}^{2}g^{2}\phi - 
\frac{3gF'\sin^{2}F}{2\pi^{2}} + \frac{2\phi}{r^{2}}
\end{eqnarray}
with the boundary conditions
\begin{equation}
  \label{eq:20}
\phi(0) =\phi(\infty)=\xi_1'(0)=\xi_1(\infty)=\xi_2'(0)=\xi_2(\infty)=0\,.  
\end{equation}
The boundary conditions for $\xi_1$ and $\xi_2$ satisfy the relation
$2\xi_1(0)+\xi_2(0)=2$. 

Finally, the effective masses of the nucleon and the $\Delta$ isobar
are expressed in terms of the hedgehog mass $M_{\mathrm{sol}}$ and the
moment of inertia $\Lambda$ 
\begin{equation}
  \label{eq:21}
M_N=M_{\mathrm{sol}}+\frac{3}{8\Lambda}\,,\qquad
M_\Delta=M_{\mathrm{sol}}+\frac{15}{8\Lambda}\,.  
\end{equation}
In the present model, the hedgehog mass is $M_{\mathrm{sol}}\simeq
1473$\,MeV and 
nucleon mass is $M_N\simeq 1562$\,MeV. Small differences between these
values and the values presented in Ref.~\cite{Meissner:1987ge} are due
to the different values of the pion mass, $m_\pi=135$\,MeV in the
present work and $m_\pi=138$\,MeV in Ref.~\cite{Meissner:1987ge}.  
\section{EMT form factors}

Using the Lagrangian in Eq.~(\ref{lag}), one can calculate
each component of the EMT as follows: 
\begin{eqnarray}
T^{{00}}\left(r\right) & = & \frac{f_{\pi}^{2}}{2}
\left(2\frac{\sin^{2}F}{r^{2}}+F'^{2}\right)
+f_{\pi}^{2}m_{\pi}^{2}\left(1-\cos F\right) \cr
&&+\frac{2f_{\pi}^{2}}{r^{2}}\left(1-\cos F+G\right)^{2}
+\frac{1}{2g^{2}r^{2}}\left\{
   2r^{2}G'^{2}+G^{2}\left(G+2\right)^{2}\right\}
\cr &&
-g^{2}f_{\pi}^{2}\omega^{2}
 -\frac{1}{2}\omega'^{2} 
+\left(\frac{3}{2}\,g\right)\frac{1}{2\pi^{2}r^{2}} \omega F'\sin^{2}F
\,,\cr  
T^{{0i}}\left(\vec{r},\vec{s}\,\right) & = &
\frac{e^{{i}lm}r^{l}s^{m}}{
  \left(\vec{s}\times\vec{r}\right)^{2}}\, \rho_J(r)\,,
\cr
T^{{ij}}\left(r\right) & = & s\left(r\right)
\left(\frac{r^{{i}}r^{{j}}}{r^{2}}
  -\frac{1}{3}\delta^{{{ij}}}\right)
+p\left(r\right)\delta^{{{ij}}}\,,   
\label{eq:emt}
\end{eqnarray}
where $T_{00}(r)$ is called the energy density. 
The vector $\vec s$ denotes the direction of the quantization axis for
the spin and coincides with the space part of the polarization vector
of the nucleon in the rest frame.  
The density of angular momentum is given by $\rho_J(r)$ while $p(r)$
and $s(r)$ are pressure and shear force densities, respectively.  
Their explicit forms are given as 
\begin{eqnarray}
\rho_{J}\left(r\right) & = &
\frac{f_{\pi}^{2}}{3\Lambda}\left[\sin^{2}F+8\sin^{4}\frac{F}{2} + 
4\sin^{2}\frac{F}{2}G-4\sin^{2}\frac{F}{2}\xi_{1}-2\xi_{1}G\right]\cr
 &  & +\frac{1}{3g^{2}r^{2}\Lambda}\left[-r^{2}\xi_{1}'G'-
\left(\xi_{1}G-G-\xi_{2}\right)\left(2G+G^{2}\right)\right]\cr
 &  & +\frac{g}{8\pi^{2}\Lambda}\Phi\sin^{2}FF'\,, \\
p\left(r\right)& = &
-\frac{1}{6}f_{\pi}^{2}\left(F'^{2}
  +2\frac{\sin^{2}F}{r^{2}}\right)-f_{\pi}^{2}m_{\pi}^{2}\left(1-\cos  
  F\right)
\cr && 
 -\frac{2}{3r^{2}}f_{\pi}^{2} \left(1-\cos
   F+G\right)^{2}+f_{\pi}^{2}g^{2}\omega^{2} \cr
 &  & +\frac{1}{6g^{2}r^{2}} \left\{
   2r^{2}G'^{2}+G^{2}\left(G+2\right)^{2}\right\}
 +\frac{1}{6}\omega'^{2}\,,\\
s\left(r\right) & = &
f_{\pi}^{2}\left(F'^{2}-\frac{\sin^{2}F}{r^{2}}\right)
-\frac{2f_{\pi}^{2}}{r^{2}}\left(1-\cos F+G\right)^{2} 
\cr &&
 +\frac{1}{g^{2}r^{2}}\left\{ r^{2}G'^{2}-G^{2}
   \left(G+2\right)^{2}\right\} -\omega'^{2}\,. 
\end{eqnarray}

The corresponding three form factors in Eq.~(\ref{Eq:EMTff}) are
finally obtained in the large $N_c$ limit as
\begin{eqnarray}
M_2(t)&=& \frac{1}{M_{\mathrm{sol}}}\int\mathrm{d}^3
        r \;T_{00}(r)\;j_0(r\sqrt{-t})-\frac{t}{5M_{\mathrm{sol}}^{2}}\,d_1(t)\,,
        \label{Eq:M2-d1-model-comp}\cr
        d_1(t)
        &=& \frac{15 M_{\mathrm{sol}}}{2}\int\mathrm{d}^3 r \;p(r)
        \;\frac{j_0(r\sqrt{-t})}{t} \,,
        \label{Eq:d1-model-comp}\cr
        J(t)
        &=& 3
    \int\mathrm{d}^3r\;\rho_J(r)\;\frac{j_1(r\sqrt{-t})}{r\sqrt{-t}}\;, 
    \label{Eq:J-model-comp}
\end{eqnarray}
where $j_0(z)$ and $j_1(z)$ represent the spherical Bessel functions
of order 0 and 1, respectively. At the zero momentum transfer $t=0$,
$M_2(0)$ and $J(0)$ are normalized as
\begin{equation}
  \label{eq:norm}
M_2(0) \;=\; \frac{1}{M_{\mathrm{sol}}}\int\mathrm{d}^3 r\;T_{00}(r) = 1
\,,\;\;\;\;\;
 J(0) \;=\; \int\mathrm{d}^3r\;\rho_J(r)=\frac12\;.
\end{equation}
These relations are very important, since the integration of $T_{00}$
should be the same as the nucleon mass, and the spin of the nucleon
should be $1/2$. 
The first condition in Eq.~(\ref{eq:norm}) is obviously seen from the
comparison of the integrand in the Hedgehog mass and the expression
for $T^{00}$. To prove the second condition in Eq.~(\ref{eq:norm}) we
integrate by part  
the terms of the bilinear combinations in derivatives
(e.g. $r^2\xi_1^{\prime2}$) in the expression of moment of inertia 
and use the equations of motion. Then the moment of inertia takes the
form 
\begin{eqnarray}
\Lambda&=&4\pi\int_0^\infty \mathrm{d}r
r^2\left\{\frac{2f_{\pi}^{2}}{3}\left(\sin^{2}F
  +8\sin^{4}\frac{F}{2}-4\sin^{2}\frac{F}{2}\xi_{1}\right) \right.
\cr &&\left.
+\frac{2}{3g^{2}r^{2}}\left\{
  \left(2-2\xi_{1}-\xi_{2}\right)G^{2}\right\}
+\frac{g}{4\pi^{2}}\phi F'\sin^{2}F\right\}\,.
\end{eqnarray}
Analogously, integrating by part the term proportional to
$r^2\xi_1'G'$ in the expression of the angular density $\rho_j(r)$,
one can show that the second condition is also satisfied.

Furthermore, the conservation of the EMT leads to the
following stability condition 
\begin{equation}
\label{eq:22}
\int_0^\infty dr\,r^2\, p(r) \;=\; 0.
\end{equation}
We can easily prove analytically that the stability
condition~(\ref{eq:22}) is satisfied within the present model.    
\begin{eqnarray}
r^{2}p\left(r\right) & = & \frac{\partial}{\partial
  r}\left[r^{3}p-2r^{3}\left(-\frac{1}{6}f_{\pi}^{2}F'^{2}
    +\frac{1}{6}\omega'^{2}\right)\right.\nonumber\\  
 &  & \quad\left.-\frac{2}{3}r^{3}\left\{ f_{\pi}^{2}g^{2}
     \omega^{2}+\frac{1}{3g^{2}r^{2}}\left\{ r^{2}G'^{2}\right\}
     -f_{\pi}^{2}m_{\pi}^{2}\left(1-\cos F\right)\right\}
 \right]\nonumber\\ 
 &  & -\frac{f_{\pi}^{2}}{3}rF' \times(\mbox{equations of
   motion}) \nonumber\\ 
 &  & -\frac{1}{3g^{2}}rG' \times(\mbox{equations of
   motion})\nonumber\\ 
 &  & -r\omega' \times(\mbox{equations of motion})\,.
\end{eqnarray}
Any reasonable model for the nucleon should satisfy
Eq.~(\ref{eq:22}). Moreover, the pressure density exhibits 
how each contribution of the mesons contribute to the shape of the
nucleon.  

\section{Results and discussion}
In this section, we now discuss the results of the EMTFFs obtained
from the $\pi$-$\rho$-$\omega$ soliton model. In Table~\ref{tab:1}, 
the relevant observables to the EMTFFs are listed in comparison with
the Skyrme model~\cite{Cebulla:2007ei} and the
$\chi$QSM~\cite{Goeke:2007fp}. Note that the Skyrme model takes the
value of the pion decay constant $f_\pi=54$ MeV such that the nucleon
mass can be fitted to the experimental data. On the other hand, the
present model and the $\chi$QSM fix it to be the experimental value
$f_\pi=93$ MeV.   
\begin{table}[ht]
\centering
\caption{
\label{tab:1}
Observables relevant to the nucleon EMT densities and 
their form factors: $\langle r_{00}^{2}\rangle$ are the mean 
square radii for the energy densities. $\langle r_{J}^{2}\rangle$
represent the squared radii of the angular momentum distribution. 
$r_{0}$ designates the position, where the sign of the pressure is
changed. $d_{1}\left(0\right)$ correspond to the
$d_{1}\left(t\right)$ form factors at the zero momentum 
transfer. $p_0(0)$ denote the pressure values at the
origin and $T_{00}(0)$ the energy densities at the origin.}

\begin{tabular}{@{}cccccccccc}
\br
Model&$\langle r_{00}^{2}\rangle$&$\langle
r_{J}^{2}\rangle$&$r_{0}$&$d_{1}(0)$&$p_0(0)$ &$T_{00}(0)$ \\ 
& [fm$^2$] & [fm$^2$] & [fm] & & [GeV/fm$^3$] & [GeV/fm$^3$] \\
\mr
$\pi\rho\omega$ soliton model &  0.78 & 0.74 &  0.55 &
-5.03&0.58 &3.56 \\
Skyrme model \cite{Cebulla:2007ei} &  0.54 & 0.92 &  0.64 &
-4.48 &0.48&2.28 \\
$\chi$QSM \cite{Goeke:2007fp} &  0.67 & 1.32 &
0.57 & -2.35&0.23&1.70 \\
\br
\end{tabular}

\end{table}
Because of this, the present work and the $\chi$QSM
overestimate the nucleon mass. The result of the $\langle
r_{00}^2\rangle$ is similar to those from the other two models, while 
that of $\langle r_J^2\rangle$ turns out to be smaller than those from 
the other models. It already indicates that the form factor $J(t)$
will fall off slower than those from the other models, which we
will discuss later.  The $D$ term, i.e. $d_1(0)$, is yielded to be
larger, compared to those of the Skyrme model and the $\chi$QSM. We
find that the values of the pressure and the energy density at the
origin are larger in comparison with the results of the other two
models.  

\begin{figure}[hbt]
\begin{center}
\includegraphics[scale=0.4]{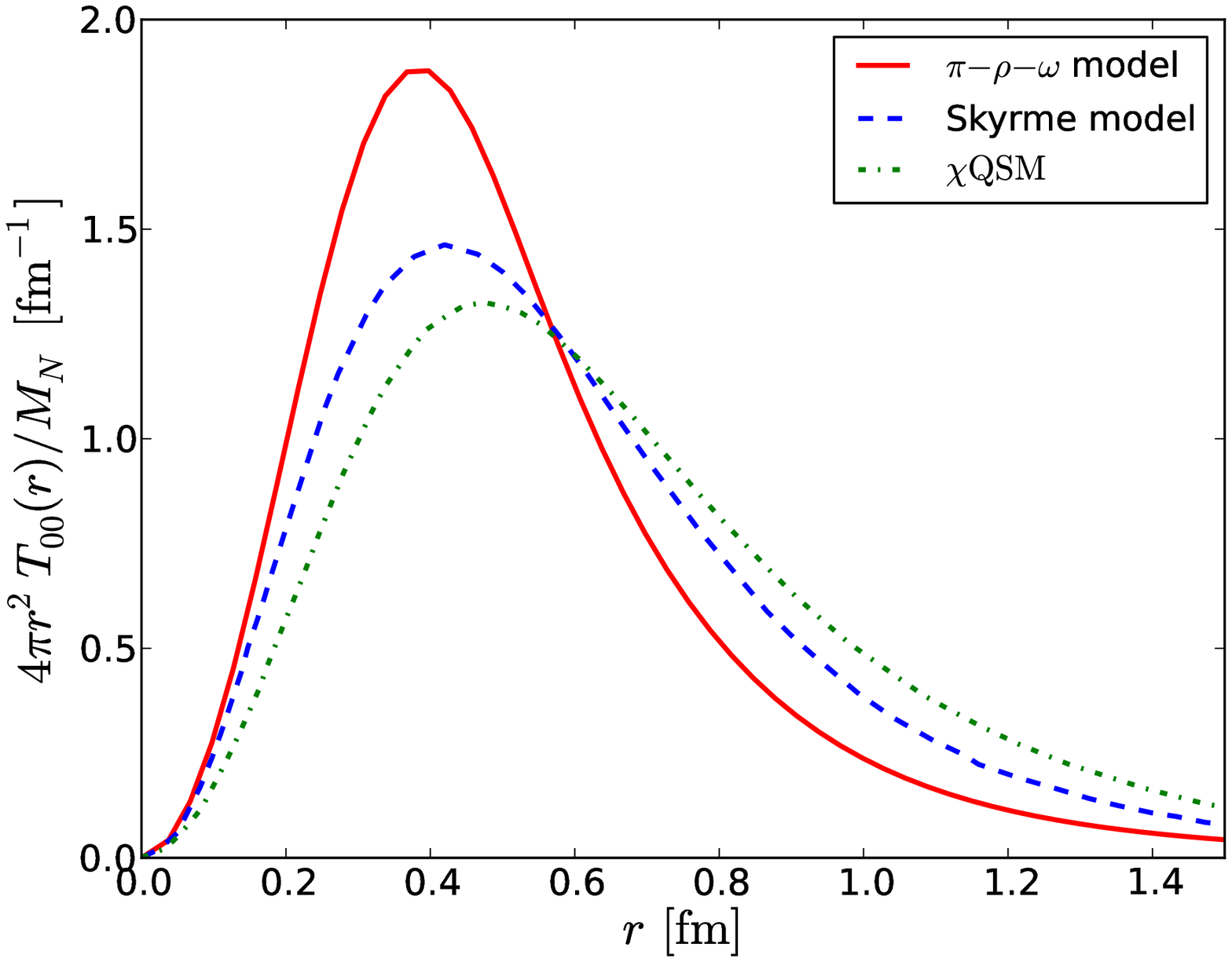}
\includegraphics[scale=0.4]{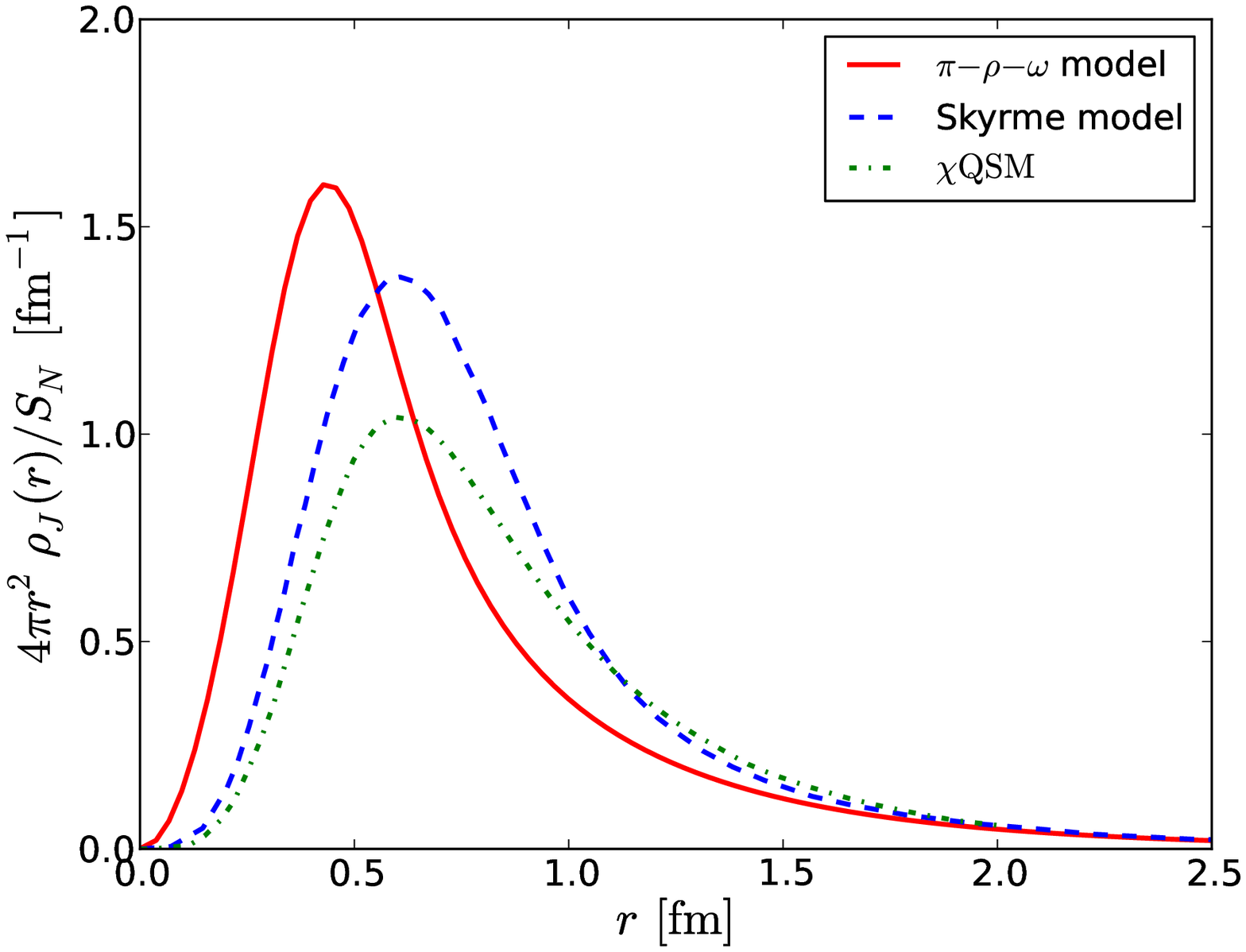}
\includegraphics[scale=0.4]{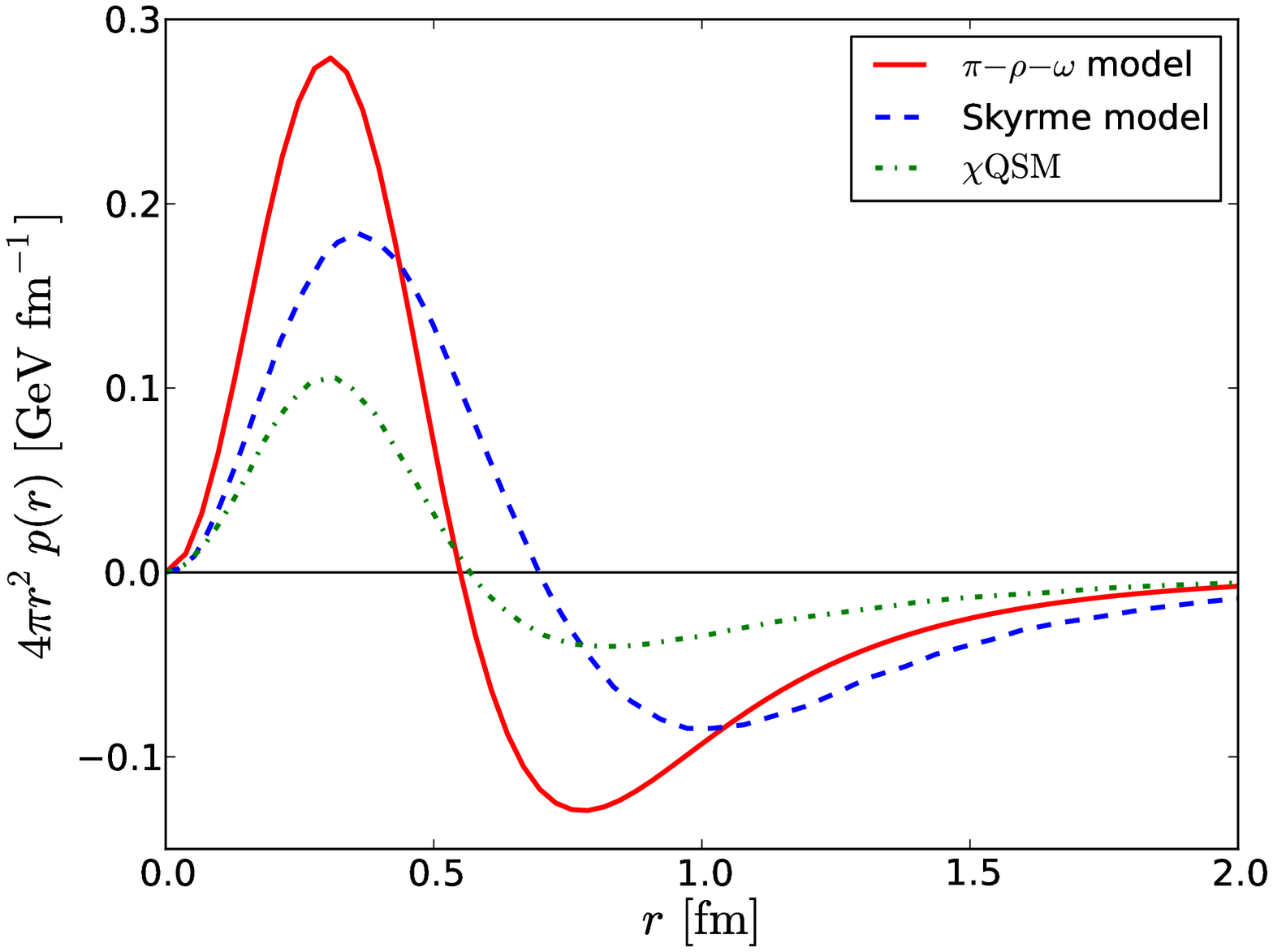}
\caption{The energy densities normalized by the nucleon mass are drawn
  in the upper panel. In the middle panel, the angular-momentum
  densities of the nucleon normalized by the nucleon spin are
  presented. The lower panel depicts the pressure densities of the
  nucleon. The solid curve represents the result of the present model,
  while the dashed and dot-dashed ones stand for those of the Skyrme
  model and the chiral quark-soliton model, respectively.     
}
\label{fig:1}
\end{center}
\end{figure}
Figure~\ref{fig:1} shows the three densities of the energy, the angular
momentum and pressure, repectively. In general, the present results
are more shifted to the center, compared with those of the Skyrme 
model and the $\chi$QSM. The pressure density becomes the most
interesting one, since it takes a picture of the nucleon internal
structure. As shown in the lower panel of Fig.~\ref{fig:1}, the
pressure density turns out to be positive in the inner part of the
nucleon but is changed to be negative as $r$ increases. However, it
should comply with the stability condition given in
Eq.~(\ref{eq:22}). So does the present result.   
 
\begin{figure}[hbt]
\begin{center}
\includegraphics[scale=0.75]{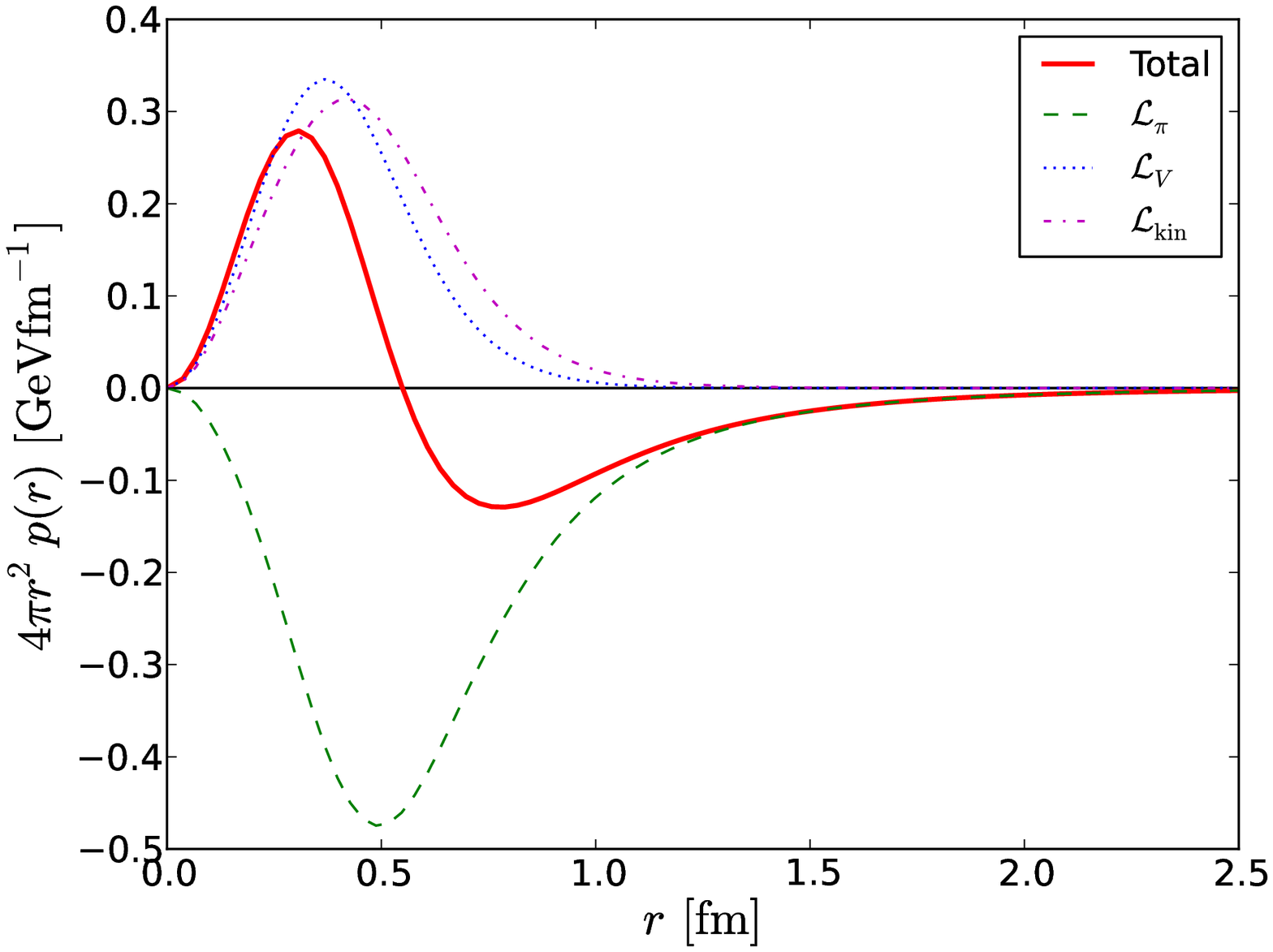}
\caption{\label{fig:2}
The pressure densities of the nucleon from each contribution. 
The dashed curve represents the contribution of the pion, while the
dotted and dot-dashed ones draw the contributions of the vector
mesons and the kinetic term, respectively. The solid curve depicts the 
total contribution.
}
\end{center}
\end{figure}
Figure~\ref{fig:2} reveals a salient feature of the pressure density. 
The pion provides a strong attraction together with the
long-range tail. As in the case of the Skyrmion, the soliton is never
stabilized with the pion only. The Skyrme term provides a repulsive
force enough to stabilize it. In the present model, the $\rho$ and
$\omega$ do play the same role as the Skyrme term. As shown in
Fig.~\ref{fig:2}, the $\rho$ and $\omega$ mesons yield repulsive
interactions, in particular, in the inner part of the nucleon. Thus,
the pressure density becomes positive in the inner part while it turns
out to be negative in the outer region with the long-range pion
tail. This feature is in line with the one-boson exchange picture of
the nucleon-nucleon interaction~\cite{Machleidt:1987hj}, as mentioned
in Introduction.

\begin{figure}[ht]
\begin{center}
\includegraphics[scale=0.4]{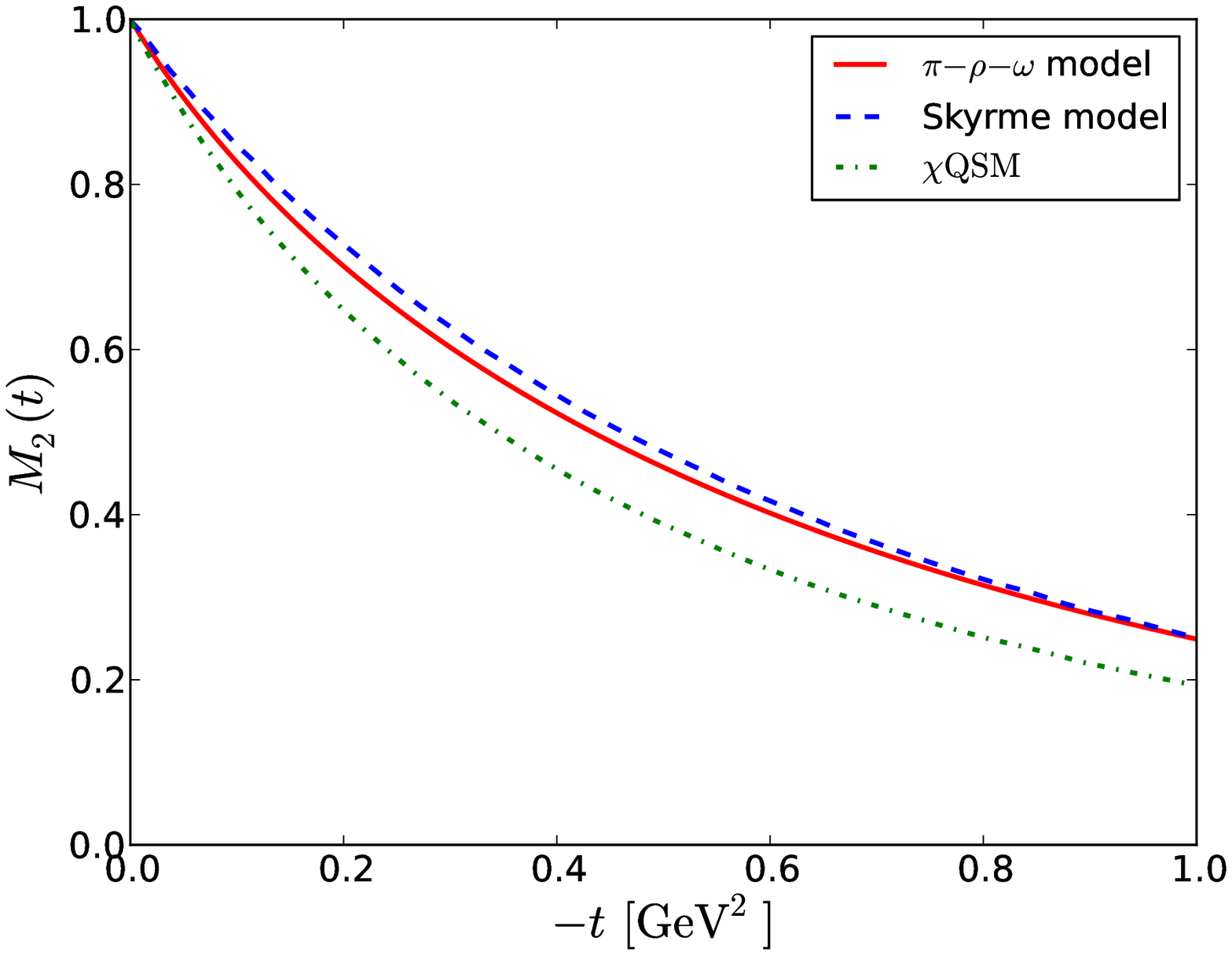}
\includegraphics[scale=0.4]{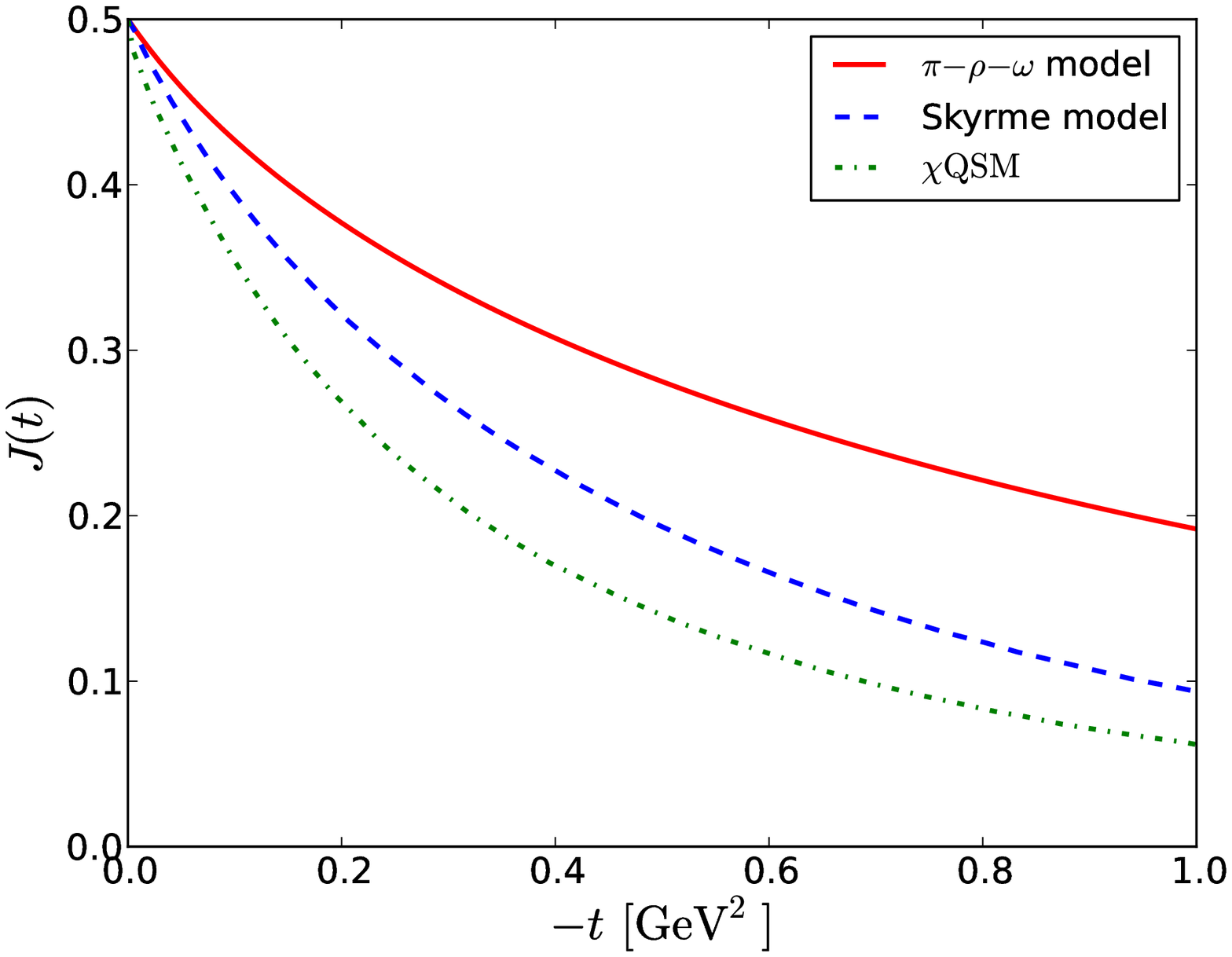}
\includegraphics[scale=0.4]{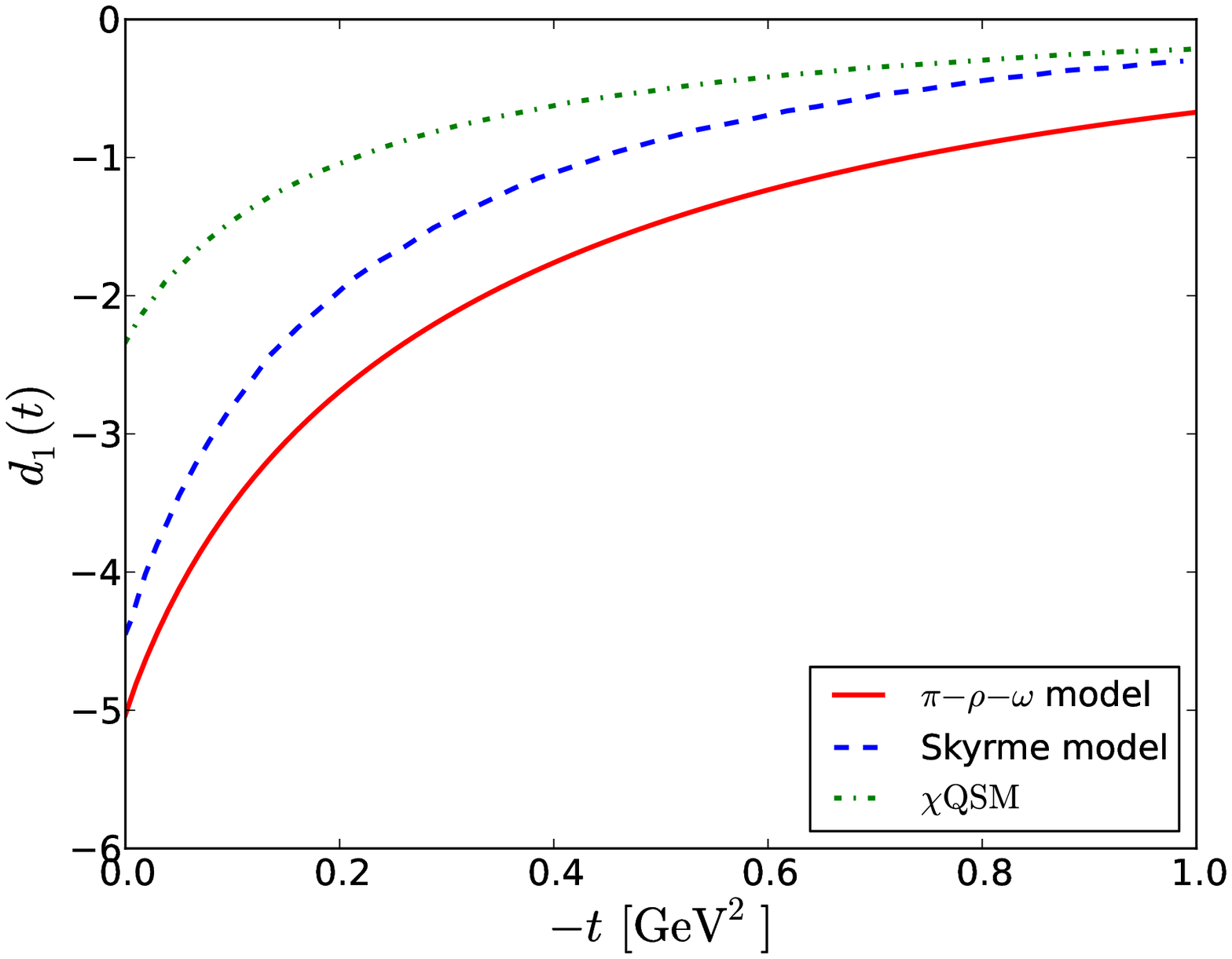}
\caption{\label{fig:3}
The dependence of form factors $M_2(t)$, $J(t)$ and $d_1(t)$
on the momentum transfer $t$. The notation is the same as in
Fig.~\ref{fig:1}.}  
\end{center}
\end{figure}
Finally, we discuss the results of the form factors in
Fig.~\ref{fig:3}. The upper panel of Fig.~\ref{fig:3} depicts 
the result of the normalized mass form factor $M_2(t)$. 
As expected from that of $\langle r_{00}^2\rangle$, the present result
shows a very similar $t$ dependence to those from the Skyrme model and
the $\chi$QSM. On the other hand, the form factor of the angular
momentum $J(t)$ and the $D$-term form factor $d_1(t)$
fall off rather slowly in comparison with those of the other two
models. 

Form factors of the proton are often parameterized with the
dipole-type form factor, $F(t) =
F(0)/(1-t/M_{\mathrm{dipole}}^2)^2$. For example, the nucleon electric
form factor is well described with this parameterization. Because of
this fact, Refs.~\cite{Cebulla:2007ei,Goeke:2007fp} fitted the EMFFFs
with the dipole-type parameterization and derived the dipole mass for
each form factor. However, we find that though the dipole-type
parameterization describes approximately well the EMTFFs, there are
still discrepancies. Thus, we use rather the $p$-pole
form factor 
\begin{equation}
  \label{eq:23}
F(t) \;=\; \frac{F(0)}{(1-t/(p M_p^2))^p},\;\;\; p\ge 1,  
\end{equation}
which parameterizes the EMTFFs quantitatively. Note that this
parameterization is often employed in lattice
QCD~\cite{Gockeler:2006zu}. Using Eq.(\ref{eq:23}), we find that the
$p$-pole mass $M_{M_2}=0.724\,\mathrm{GeV}$ for the mass form factor
$M_2(t)$ with $p=2.17$, $M_{J}=0.786\,\mathrm{GeV}$ for $J(t)$ with
$p\approx 1$, and $M_{d_1}=0.510\,\mathrm{GeV}$ for $d_1(t)$ with
$p=1.57$. Note that the $p$-pole parametrization is defined for
$p\ge 1$ to satisfy an analytic behavior at $t=0$. Because of this,
the value of $p$ for $J(t)$ is approximately fitted to $p\approx 1$.       

\section{Summary and outlook}
In the present work, we aimed at investigating the energy-momentum
tensor form factors of the nucleon, based on the $\pi$-$\rho$-$\omega$
soliton model. Having fixed all the relevant parameters in the mesonic
sector, we were able to derive the densities for the form factors. 
We discussed the results of the densities in comparison with the two
different solitonic model, i.e., the Skyrme model and the chiral
quark-soliton model. The results were in general more shifted to the
inner part of the nucleon, compared with these two models. The present
model was shown to satisfy the stability condition. the result of the
pressure density exhibited each role of the $\pi$ and the vector
mesons: While the pion provides the strong attraction, the $\rho$ and
$\omega$ yield the repulsive force that balances in such a way that
the stability condition is satisfied. We finally discussed the results
of the three form factors: the mass form factor, the angular-momentum
form factor, and the $D$-term formfactor. While that of the mass form
factor was quite similar to those of the other models, the results of
the angular-momentum and $D$-term form factors turned out to fall off
more slowly than those of the Skyrme model and the chiral
quark-soliton model. 

The energy-momentum tensor form factors provide important information
on how the nucleon undergoes changes in nuclear
matter~\cite{Kim:2012ts}. It is in particular of great interest to
study them within the present framework, since the medium-modified
$\pi$-$\rho$-$\omega$ soliton model connects the change of the vector
meson in nuclear matter with the medium modification of the
nucleon~\cite{Jung:2012sy}. Thus, the energy-momentum tensor
form factors in nuclear matter within the medium-modified
$\pi$-$\rho$-$\omega$ soliton model will shed light on the physics of
the nucleon in medium. The corresponding investigation is under
way~\cite{Jung:2013}. Last but least, it is also interesting to
examine the quantum corrections, which might be of great importance
in the context of the energy-momentum tensor form
factors~\cite{Meier:1996ng}. This can be considered as a future work.

\ack
One of the authors (H.-Ch.K) is grateful to M. V. Polyakov for
suggesting this work. He is also thankful to P. Schweitzer for
discussions and comments. The present work is supported by Basic
Science Research Program through the National Research Foundation
(NRF) of Korea funded by the Korean government (Ministry of Education,
Science and Technology -- MEST), Grant Number: 2012-0008469 
(J.-H.J. and U.Y) and Grant Number: 2012004024 (H.Ch.K.).

\end{document}